\documentclass[11pt]{article} 
\usepackage{moriond,epsfig} 

\def\PRL{\em Phys. Rev. Lett.}

\def\be{\begin{equation}} 
\def\ee{\end{equation}} 
\def\bea{\begin{eqnarray}} 
\def\eea{\end{eqnarray}}

\def\etal{{\it et al.}}

\begin{document} 
\vspace*{4cm} 
\title{
CLEO Results on Upsilons \footnote{ Results reported at the XXXIXth Rencontres de Moriond, QCD and Hadronic Interactions, La Thuile, Italy. } }
 
\author{ J.E. Duboscq \\
 ( for the CLEO Collaboration ) } 
 
\address{
Laboratory for Elementary-Particle Physics, Cornell University, Ithaca NY 14853, USA
} 
 
\maketitle\abstracts{ 
I present highlights of recent CLEO  studies on the $\Upsilon (1S)$, $\Upsilon (2S)$, and
$\Upsilon (3S)$. The preliminary results presented here include the first observation
 of a hadronic cascade in the $\Upsilon$ system not involving  pions,  $\Upsilon (3S) \to \gamma \chi_{b1,2} (2P)$ followed by    $\chi_{b1,2} (2P) \to \omega \Upsilon (1S)$.  I also 
 present preliminary results on a search for inclusive $\psi$ production in $\Upsilon(1S)$ decays. I end with a search for nine different two  body hadronic decays of the $\Upsilon(nS)$, $n=1,2,3$, resulting in the discovery of two such decays, and vastly improved upper limits on  
 the branching ratios of the rest.
} 
 
\section{The CLEO Detector and CLEO Data}

    During the year 2002, CLEO took data at the $\Upsilon (1S)$, $\Upsilon (2S)$ and
     $\Upsilon (3S)$, resulting in approximately 20 million hadronic events on the $\Upsilon (1S)$,
      10 million hadronic events on the $\Upsilon (2S)$ and 5 million hadronic events on the $\Upsilon (3S)$. In addition to data taken on the peaks of each resonance, data were also taken below the resonances for background purposes and in scans across the resonances.
    
 The data used in the following studies were taken with the CLEO3 detector at the CESR
 $e^+e^-$ storage ring. The detector includes a silicon microvertex detector and a drift
  chamber, as well as a crystal calorimeter and ring imaging cerenkov (RICH) detector for
   hadron ID. Muon chambers surround the detector. The tracking volume is in a uniform
    1.5 T magnetic field. 
    
 \section{ What are Upsilons ? }
 The $\Upsilon$ resonances are bound states of $b$ and $\overline{b}$ quarks. The spectrum of
 bottomonium is very much like that seen in positronium. The $\Upsilon (4S) $ state is the source 
of B mesons at the B factories. The states in question in this talk are all below $B \overline{B}$ 
threshold. The $\Upsilon(nS)$ states are produced directly in $e^+e^-$ collisions. The 
other bottomonia states are  reached by cascades from these resonances. The relative rates 
of cascade decays are a particularly important testing ground for Lattice QCD - 
should these rates be accurately predicted,  our overall confidence in the use of LQCD 
in other arenas will be bolstered.

\section{ Observation of $\chi_{b1,2} (2P) \to \omega \Upsilon (1S)$ }
 Up to now, the only transitions observed among  bottomonia states have involved the 
emission of either photons or pion pairs. In this search \cite{chib} we looked for the transition 
 sequence  $\Upsilon (3S) \to \gamma \chi_{b1,2} (2P)$ followed by 
    $\chi_{b1,2} (2P) \to \omega \Upsilon (1S)$.
  The search  involves looking for 2 leptons from the $\Upsilon(1S)$ decay recoiling against a 
  photon and $\pi^+\pi^-\pi^0$ in data taken at the $\Upsilon(3S)$ resonance. Of note is that no
   hadron particle identification is needed for this analysis, as decays with Kaons are kinematically 
   forbidden. Also, the mass spectrum in this sample for the dilepton mass from the $\Upsilon(1S)$ is 
   so clean that no lepton ID cuts are required either. The photon is required to have energy 
   between 50 and 250 MeV. The above cuts give a very clean $\omega$ peak in the 3 pion mass 
   spectrum. Plotting the dilepton mass against the $\omega \gamma$ recoil mass reveals a very clean signal. 
   The energy cut on the photons exclude the presence of the $\eta_b(3S)$ leaving the signal 
   channel as the only possible decay mode observed.
  In order to extract the relative contributions of the $\chi_{b1}$ and $\chi_{b2}$ states, we fit the 
  photon energy spectrum, and obtain:
  \begin{eqnarray}
  BR ( \chi_{b1} (2P) \to \omega \Upsilon(1S) ) & = & 1.63 ^{+0.35+0.16}_{-0.31-0.15} \% \\
   BR ( \chi_{b2} (2P) \to \omega \Upsilon(1S) )  &= & 1.10 ^{+0.32+0.11}_{-0.28-0.10} \% 
   \end{eqnarray}
   
   This is the first observation of a non-pion hadronic transition in the bottomonium system.

   \section{Charm Production in $\Upsilon(1S)$ Decays }
   The $\Upsilon$ system offers a glue rich environment in which it is interesting to study the mechanisms of charm production. The belief is that charm should be produced through the color octet mechanism.
   Our study focused on searching for the inclusive production of the $\psi$ in the decays of the $\Upsilon(1S)$. To do this we searched for $\psi \to e^+e^-, \mu^+\mu^-$ decays in the data taken at the $\Upsilon (1S) $ resonance. We subtracted the luminosity and beam energy scaled contribution from continuum processes using data taken below the resonance.
   Additional cuts were used to suppress radiative returns to the $\psi$ and $\psi(2S)$. As a cross check, the same analysis method was used on $\Upsilon (4S)$ data to verify that its results were as expected. 
    The resulting preliminary branching fraction is found to be
    \begin{equation}
    Br ( \Upsilon (1S) \to \psi X) = (6.1 \pm 0.3 \pm 0.6 ) \times 10^{-4}
    \end{equation}
 Also of interest is the beam energy scaled momentum spectrum of the $\psi$'s in this process. It appears that this spectrum is softer  than would be expected from a naive color octet model, although it might be possible to address this issue with the emission of soft gluons in the theoretical calculations.
 
 \section{ Two Body Hadronic Decays of the $\Upsilon$ }
 To date, no purely hadronic decays of the $\Upsilon (nS) (n=1,2,3) $ have been observed.
 In addition, it is of interest to see if these decays in the $\Upsilon$ system will follow what has 
 been called variously the $14 \%$ or $12\%$ rule seen in $\psi$ and $\psi'$ decay. This rule of thumb
  follows from the naive observation that in $\psi (nS)$ decay, the relative production widthes of the 
  $e^+ e^-$ and $q \overline{q}$ final states should be proportional. Extending this ansatz to the
   fully hadronised final states leads to the expectation that the branching ratio of $\psi' \to X Y$ 
   should be  $12\%$ of the branching ratio of $\psi \to X Y $. This rule is grossly obeyed in the 
 $\psi$   system, except that the decay rate $\psi' \to \rho \pi$ seems too small. This is occasionally referred 
    to as the $\rho \pi$ puzzle. 
    In the $\Upsilon$ system, with 3 states below B threshold, the rule  translates to a 48\% (72\%) rule when comparing 2S (3S) to 1S decays.
    
    This analysis concentrates on a search for $\Upsilon$ decays to those final states
     that have large rates in $\psi$ decays. The modes chosen are
        $\rho \pi$ ,
$K^*(892) \overline{K}$,
$\rho  a_2 (1320)$, 
$ \omega f_2 (1270)$,
$ \phi f_2'(1525)$,
$K^*(892) \overline{K}_2^* (1430)$,
$b_1(1235) \pi$,
$K_1(1270) \overline{K}$, and 
$K_1(1400) \overline{K}$.

\begin{figure}
\begin{center}
\psfig{figure=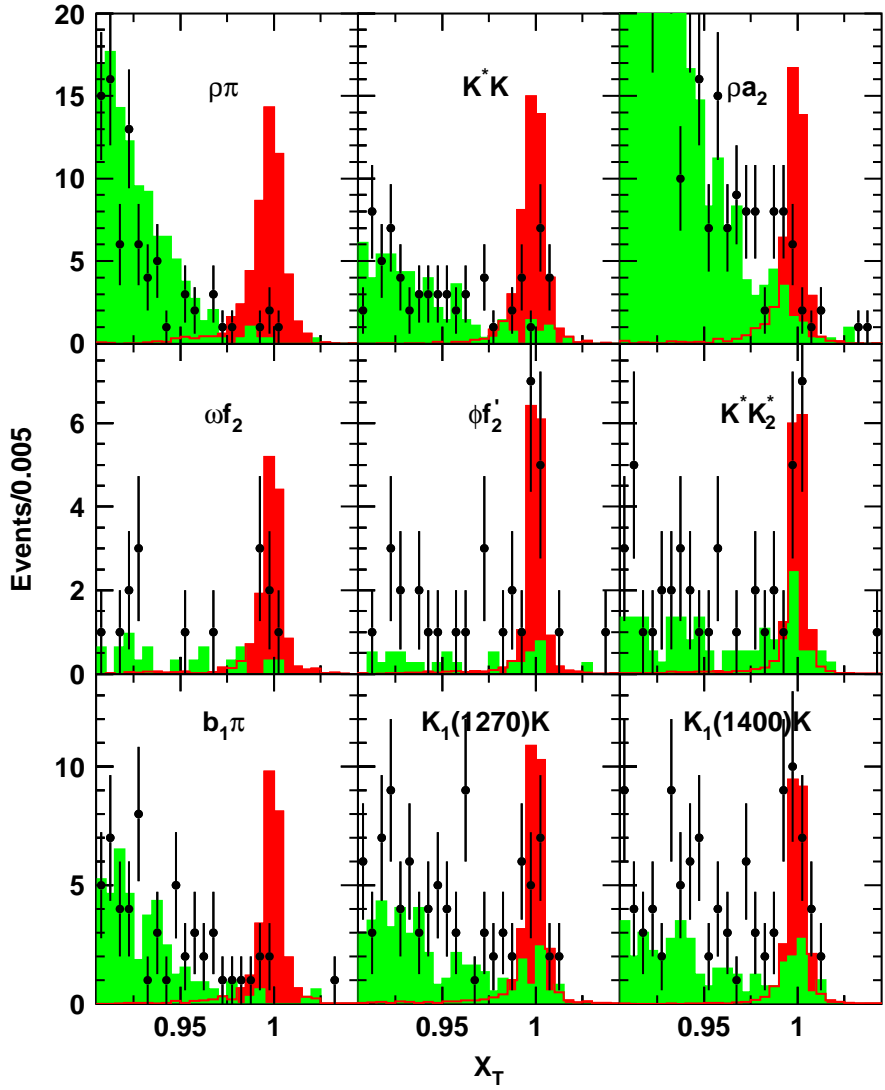,height=6in}
\caption{Scaled Total Energy For 2 Body Hadronic Decays of the $\Upsilon(1S)$. The points 
with error bars are data from the $\Upsilon(1S)$, while the light histogram is the scaled 
expectation from non $\Upsilon(1S)$ data, and the dark histogram represents arbitrarily 
normalized signal Monte Carlo. 
\label{fig:upsxy}}
\end{center}
\end{figure}

        Fig~\ref{fig:upsxy} shows the scaled total energy for the decay products of 
 the $\Upsilon(1S)$ decays. Note that there
         are statistically significant signals for the $\phi f_2'(1525)$ and $K_1(1400) \overline{K}$
          final state, leading to our claim of first observation for these modes. All other final states lead
           to vastly improved upper limits on the branching ratios in the range of 
  $10^{-5}$ to $10^{-6}$, as summarized in Table~\ref{tab:exp}. These limits indicate that 
 branching fractions to these fully
             hadronic states are suppressed by a factor of more than 100 relative to those seen in the
              $\psi$ system. Unfortunately, this means that much more data, and much more work will be
               needed to test the $12\% $ rule in the $\Upsilon$ system.

\begin{table}[t]
\caption{ Preliminary Results on Branching Ratios for $\Upsilon$ Decays to Two Body Hadronic States - all results are in units of $10^{-6}$ - limits are 90 $\%$ confidence level upper limits.\label{tab:exp} }
\vspace{0.4cm}
\begin{center}
\begin{tabular}{|c|c|c|c|}
\hline
Final State   &  $\Upsilon(1S)$ & $\Upsilon(2S) $& $\Upsilon(3S) $ \\ \hline \hline
$\rho \pi$ &  $<4 $ & $<11$ & $<22 $ \\ \hline
$K^*(892) \overline{K}$ & $<11$ & $<8$ & $<14$ \\ \hline
$\rho  a_2 (1320)$ & $<19$ & $< 24 $ & $<30$ \\ \hline
$ \omega f_2 (1270)$  & $<7$ & $< 11 $& $< 8 $ \\ \hline
$ \phi f_2'(1525)$ & $ 7^{+3}_{-2}\pm1$ & $<17 $ & $< 14 $ \\ \hline
$K^*(892) \overline{K}_2^* (1430)$ & $<19 $ & $<32 $ & $ < 28 $ \\ \hline
$b_1(1235) \pi$ & $ <8 $ & $<12$ & $<18 $ \\ \hline
$K_1(1270) \overline{K}$ & $8$& $<11$ & $<17 $ \\ \hline
$K_1(1400) \overline{K}$ & $14^{+3}_{-2}\pm2 $&  $< 33 $ & $<22$ \\ \hline
\hline
\end{tabular}
\end{center}
\end{table}

\section{Conclusions}
         
         CLEO is now fully exploiting the world's largest sample of $\Upsilon$ decays. We have reported on the observation of $\chi_{b1,2} \to \omega \Upsilon(1S)$, which the first
          observation of a transition in the $\Upsilon$ system not involving a pion or photon.
          We have also measured the rate for inclusive $\psi$ production in $\Upsilon(1S)$ decays,
 and measured the momentum spectrum of the $\psi$'s. This will hopefully be useful in 
 shedding light
 on the mechanisms of gluonic production of $c \overline{c}$ pairs. We have also presented 
 the first observation of two hadronic decay modes in the $\Upsilon (1S) $ system and 
 set stringent limits
 on 7 others from the $\Upsilon(1S)$ and 9 each from the $\Upsilon (2S)$ and $\Upsilon(3S)$.       All results reported here are preliminary.

\section*{Acknowledgments} 
I would like to thank the conference organizers for this lovely speaking invitation,
 and would also like to extend my appreciation to the very courteous hotel staff.  
 I would also like to thank the National Science Foundation for its support over the years
 and my colleagues at CLEO and CESR for their hard work.

\end{document}